# Switching of a single ferromagnetic layer driven by spin Hall effect

G. Finocchio,[1] M. Carpentieri,[2] E. Martinez,[3] B. Azzerboni[1]

[1] Department of Electronic Engineering, Industrial Chemistry and Engineering. University of Messina, C.da di Dio, I-98166, Messina, Italy.

[2] Department of Electrical and Information Engineering, Politecnico of Bari, via E. Orabona 4, I-70125 Bari, Italy.

[3] Universidad de Salamanca. Plaza de los Caidos s/n, E-38008, Salamanca. Spain.

**Abstract:** The magnetization switching of a thin ferromagnetic layer placed on top of a heavy metal (such as Pt, Ta or W) driven by an in-plane current has been observed in recent experiments. The magnetization dynamics of these processes is studied in a full micromagnetic framework which takes into account the transfer-torque from spin Hall effect due to the spin-orbit coupling. Simulations indicate that the reversal occurs via nucleation of complex magnetization patterns. In particular, magnetic bubbles appear during the reversal of the magnetization in the perpendicular configuration while for the in-plane configuration, nucleation of vortexes are observed.

PACS: 85.75.-d

The switching of a single ferromagnet driven by an in-plane bias current was recently observed in experiments on bi-layer composed by a ferromagnet/heavy-metal (F/H) with large spin-orbit coupling. This discovery has given rise to potential new routes in the implementation of high speed and low power storage applications.[1,2,3] In particular, in a thin film geometry (F/H), such as the one shown in Fig. 1(a) or (b),(see also of Fig. 1 of ref.[1] and Fig. 3a of ref.[2]) a charge current flowing in the in-plane direction is able to reverse the magnetic state of the ferromagnet. The mechanism which is responsible for the switching was subject of debate. In the pioneering experiment,[1] the magnetization switching was attributed to the Rashba field (RF) due to the strong Rashba spin-orbit coupling, indeed, the existence of a spin-transfer torque of the Slonczewski-type was also conjected in that experimental framework. However, in the experimental results achieved in[2,3,4] the switching was attributed to the spin-transfer torque from the spin-Hall effect[5,6] (ST-SHE), being the measured RF negligible in those latter devices. In particular, the comparison between the experimental data on different heavy-metal such as Pt,[3] Ta,[2] and W[4] clearly indicates that the ST-SHE is the main mechanism which drives the switching. In fact, this conclusion derives from the experimental results that the spin-Hall angle in Pt has opposite sign compared to the one in Ta or W and consequently the electric current to achieve the switching in presence of Pt has opposite sign than the one used in presence of Ta or W. On the other hand, the sign of the RF torque should be independent on the heavy-metal used. For a complete review and an analysis of the experimental measurements of the spin-Hall effect see Refs.[7] and [8]. However, we stress the fact that the experimental findings of the high value of the spin-Hall angle ($\alpha_H = -0.3$) in bilayer W/CoFe (at least three times larger than the one in bilayer with Pt) is, from a technological point of view, an important step towards the realization of a different scheme of non-volatile magnetic solid state memory[2,9] and nanoscale auto-oscillators[10]. In addition, we point out that recent experimental measures of the magnetization self-oscillations[10,11] clearly indicate the presence of a negative damping mechanism (Slonczewski-type torque from SHE) which compensates the natural magnetic losses (positive damping) of the magnetization.

With this in mind, here we present the results of numerical computations of the magnetization reversal processes driven by the ST-SHE, demonstrating that the exclusive effect of the ST-SHE is able to efficiently switch the magnetic state of a nanomagnet.

In a pioneering paper,[3] the experimental features have been qualitatively studied by means of a modified Stoner-Wohlfarth including the effects of the non-uniformities and the thermal fluctuations. Our full micromagnetic computations show that in both systems studied experimentally[1,2] the switching occurs via a complex nucleation process. In the Co perpendicular ferromagnet, the nucleation of magnetic bubbles is observed, while in the CoFe in-plane ferromagnet the nucleation of magnetic vortexes is achieved.

The ST-SHE is included in the Landau-Lifshits-Gilbert (LLG) equation as additional torque of Slonczewski-type[3, 12]

$$\frac{d\mathbf{M}}{dt} = -\gamma_0 \mathbf{M} \times \mathbf{H_{EFF}} + \frac{\alpha}{M_S} \mathbf{M} \times \frac{d\mathbf{M}}{dt} - \frac{d_J}{M_S} \mathbf{M} \times \mathbf{M} \times \boldsymbol{\sigma}$$ (1)

where $\mathbf{M}$ and $\mathbf{H_{EFF}}$ are the magnetization and the effective field. The $\mathbf{H_{EFF}}$ takes into account the standard micromagnetic contributions, the exchange, the magnetocrystalline anisotropy, the external, and the self-magnetostatic fields. We introduced a Cartesian coordinate system where the direction of the current flow is the *x*-direction while the *z*-direction coincides with the out-of-plane axis. The coefficient $d_J$ is given by $d_J = \frac{\mu_B \alpha_H}{e M_S d} j$, being $j$ the charge current density, $\mu_B$ the Bohr Magneton, $e$ the electric charge and $d$ the thickness of the Co-layer, $\alpha$ is the Gilbert damping, $M_S$ is the saturation magnetization, and $\gamma_0$ is the gyromagnetic ratio. $\boldsymbol{\sigma}$ is the direction of the spin current in the Pt (*y*-direction in our Cartesian coordinate system). $\alpha_H$ is the spin hall angle given by the ratio between the amplitude of transverse spin current density generated in the Pt and the charge current density flowing in it. The equation (1) can be also rewritten to identify the contribution of a

field like torque term proportional to the Gilbert damping, and by considering the dimensionless form, it is:

$$\frac{d\mathbf{m}}{\gamma_0 M_S dt} = -\frac{1}{(1+\alpha^2)}\mathbf{m}\times\mathbf{h}_{EFF} - \frac{\alpha}{(1+\alpha^2)}\mathbf{m}\times\mathbf{m}\times\mathbf{h}_{EFF}$$
$$-\frac{d_J}{(1+\alpha^2)\gamma_0 M_S}\mathbf{m}\times\mathbf{m}\times\boldsymbol{\sigma} + \frac{\alpha d_J}{(1+\alpha^2)\gamma_0 M_S}\mathbf{m}\times\boldsymbol{\sigma}$$

(2)

being $\mathbf{m} = \mathbf{M}/M_S$ and $\mathbf{h}_{EFF} = \mathbf{H}_{EFF}/M_S$.

The first simulated system is shown in Fig. 1(a) and it is similar to the one measured in the experimental framework presented in Ref. [1], a square Co layer (500 nm in side) of 0.6 nm in thickness coupled with a 3 nm in thickness of Pt. The parameters used for the simulations are: exchange constant $A=2.0\times10^{-11}$ J/m, $\alpha=0.1$, $M_S=900\times10^3$ A/m, perpendicular anisotropy constant $k_U=1.1\times10^6$ J/m$^3$, $\alpha_H=0.08$.[13] The magnetization switching is studied in presence of an in-plane bias field parallel to the direction of the electric current flow and for a current pulse of 10 ns. The simulations have been performed in presence of thermal fluctuations and, if not specified, the temperature is $T=300$ K. The maximum applied field is 60 mT, smaller than the field necessary to set in-plane the Cobalt magnetization.

We identified three different regions of behavior as function of the bias field and current: (i) *direct switching* (*DS*), (ii) *switching via an intermediate state* (*IS*), and (iii) *stochastic switching* (*SS*). Fig. 2 summarizes a phase diagram of the switching of the average normalized *z*-component of the magnetization <$m_Z$> from positive to negative as function of the applied field and current ($H>0$ mT, $J>0.5\times10^8$ A/cm$^2$).

The *direct switching* regime is observed for magnetic field larger than 10 mT. The magnetization reversal is achieved before the current pulse is removed. The negative to positive switching is achieved by only reversing the sign of the current and maintaining the same bias field. Fig. 3(a) summarizes the switching processes, <$m_Z$> vs time, for different fields and current densities (the values for each process can be read directly in the figure). In order to deeply understand the switching mechanism, the time evolution of the spatial distribution of the magnetization has been also analyzed. Our results indicate that the switching occurs via a complex nucleation process. Firstly, the nucleation of one or more magnetic bubbles is achieved giving rise to regions with a reversed magnetization , [14] and finally those magnetic bubbles expand in the whole cross section completing the reversal.[15] The spatial position where bubbles are nucleated depends on the seed of the thermal field, in other words the place where the magnetization is locally reversed changes for each different simulation realization, as expected. A systematic study of the magnetization dynamics as function of the temperature shows the important role of the thermal field. Fig. 3(b) displays the switching processes for $H$=50 mT and $J$=1.0x10$^8$ A/cm$^2$ at different temperatures ($T$=200, 220, 250, and 280 K), the main result is the presence of a critical temperature below which the switching does not occur. This means that the switching driven by ST-SHE is thermally activated at least for the geometrical-physical parameters and the range of current densities considered in this numerical experiment.

An intricate scenario is indeed observed at large current densities (field larger than 10 mT and $J$>1.5x10$^8$ A/cm$^2$), the *switching* occurs *via an intermediate state* where the current drives the magnetic configuration in a complex intermediate state that evolves toward the reversed magnetic state after the current pulse is switched off. Fig. 4(a) summarizes the switching processes for different fields and current densities. An example of the magnetic configuration in the intermediate regime ($H$=25 mT, $J$=2.0x10$^8$ A/cm$^2$, point A) is plotted in Fig. 5 (the color is related to the *z*-component of the magnetization, blue positive and red negative, the arrows indicate the in-plane component of the magnetization). The presence of this intermediate state can be seen experimentally looking at the anomalous Hall resistance without removing the current.[1] The origin of those intermediate states is related to a trade off among the ST-SHE, the exchange and the self-magnetostatic fields. An increasing of the exchange energy (larger values of *A*) implies the application of larger current to nucleate those intermediate states. On the other hand, the effect of the saturation magnetization is non-trivial because an increasing/decreasing of $M_S$ reduces/amplifies the effect of the ST-SHE term and increases/decreases the out of-plane component of the demagnetizing field. A complete analysis on this behavior will be presented elsewhere.

The *stochastic switching* regime is observed for magnetic field smaller than 10 mT. A large enough current drives the magnetic configuration in a multidomain state characterized by a zero value of the $<m_Z>$ (see Fig. 4(b)). After the current pulse is switched off, the magnetic configuration can evolve toward the full magnetization reversal or the previous initial state. For instance, Fig. 4(b) shows an example of the two possible cases described above, compare the curves $J$=1.5 and 3.0 x$10^8$ A/cm$^2$.

We underline that, while in the switching via an intermediate state the magnetization reversal always occurs after the current density is switched off, differently in the *stochastic switching* the magnetization reversal can be observed or not after the current density is switched-off. At the same way, for a given switching process (same current and field), the magnetization reversal can be achieved or not depending on the seed of the thermal fluctuations.

The second simulated system is a three terminal device composed by a Magnetic Tunnel Junction (MTJ) CoFe(2)/MgO(1)/CoFeB(4) (the thicknesses are in nm) deposited on the top of a Ta extended film with a thickness of 6 nm similar to the one measured in the experimental framework by Liu *et al*[2]. The MTJ has an elliptical cross section of 300 nm x 100 nm and it is realized in order to have its easy axis perpendicular to the in-plane current flow (see Fig. 1(b)). The parameters used for the simulations are: exchange constant $A$=2.0x$10^{-11}$ J/m, $\alpha$=0.015, $M_S$=1000x$10^3$ A/m, $\alpha_H$=-0.15. Due to the different sign of the spin-Hall coefficient in CoFeB/Ta with respect to Co/Pt, the switching is achieved with the opposite sign of the current. For $T$=0 K, the switching scenario is similar to that already observed for switching processes driven by spin-polarized current in collinear spin-valves.[16, 17] The switching occurs via a nucleation process with the presence of complex magnetic pattern including also magnetic vortex states. Fig. 6(a) shows the time evolution of the average normalized magnetization for the reversal from parallel (P) to antiparallel (AP) (the reference layer is the CoFeB(4) of the MTJ) achieved for $H=0$ mT and $J=9\times10^7 \text{A/cm}^2$. The insets display some snapshots of the spatial distribution of the magnetization during the switching process related to the points A, B, and C as indicated in the figure. Fig. 6(b) and (c) show the switching time without and with the thermal fluctuations ($T=300$ K) respectively. The switching time, computed by averaging over 50 different stochastic realizations for each value of the current, is reduced in presence of thermal fluctuations at $T=300$ K, and the process is thermally assisted. An interesting additional aspect is the possibility to achieve fast switching (writing time smaller than 5 ns) for current densities close to $10^7$ A/cm$^2$, see for example Fig. 6(c). The switching is also thermally activated being the critical current density smaller in presence of thermal fluctuations ($J_C=-6.7\ 10^7$ at $T=0$ K and $J_C=-2\ 10^7$ at $T=300$ K).

In presence of a low field applied along the in-plane hard axis of the ellipse, we find that the switching of the magnetization is recovered with a quasi-coherent rotation, similarly to the results reported in Ref.[18]. As example, the switching process P→AP achieved for $H=5$ mT and $J=4.25\times10^7$ A/cm$^2$ is displayed in Fig. 7(a) (time evolution of the average normalized three component of the magnetization). The snapshots included as insets clearly indicate a switching without domain nucleation. We summarized the switching time ($T=0$ K) as function of the current density ($T=0$ K, $H=5$ mT) in Fig. 7(b). Our results underline that the presence of a low bias magnetic field can optimize the switching process such as reduce the critical current and the switching time compared to the solution at zero bias field.

In this three terminal device the bit information can be written with a current that does not flow in the cell memory. This aspect gives rise to the advantage, compared to the standard spin-transfer-torque MRAM where to write the bit the current flows in the memory cell, of increasing the durability of the memory cell by reducing the possibility of the electric breakdown of the MTJ tunnel barrier. Micromagnetic simulations also predict that, biasing the MTJ with a subcritical current density via the third terminal, the in-plane current density to achieve the fast switching can be reduced by approximately an order of magnitude (close to $5 \times 10^6$ A/cm$^2$).

In summary, we have studied the magnetization dynamics in systems composed by a heavy metal coupled with a ferromagnet in presence of an in-plane charge current. Full micromagnetic simulations demonstrated that the switching can be achieved by means of the ST-SHE only, either for perpendicular and in-plane ferromagnets, and it occurs via a nucleation process. The perspective opened by our results is very promising to predict and define the best condition to improve the device efficiency, reducing the current densities needed to obtain STT-SHE switching.

The authors would like to thank D. C. Ralph for a critical reading of the manuscript and for the suggestions on how to implement the model. The authors thank Luqiao Liu for helpful discussions. This work was supported by project MAT2011-28532-C03-01 from Spanish government, project SA163A12 from Junta de Castilla y Leon, and project PRIN2010ECA8P3 from Italian MIUR.

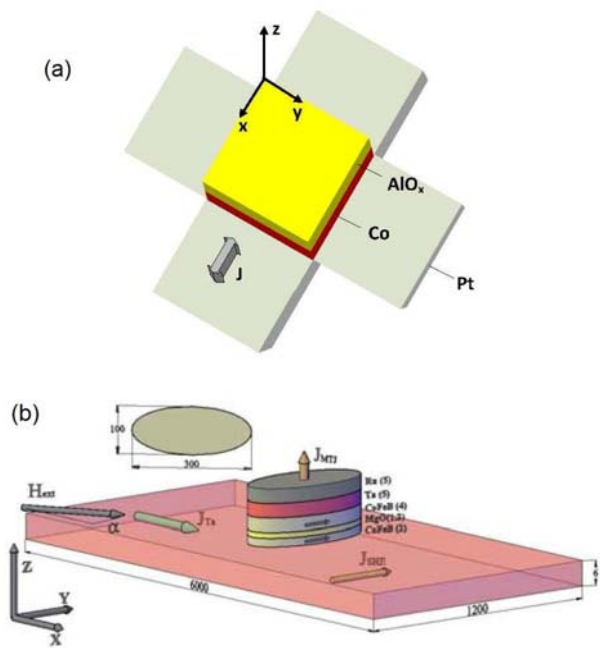

FIG. 1 Sketch of the device geometries. (a) square Co layer of 0.6 nm in thickness (500 nm in side) coupled with a 3 nm in thickness of Pt and (b) Magnetic Tunnel Junction (MTJ) CoFe(2)/MgO(1)/CoFeB(4) (the thicknesses are in nm) deposited on the top of a Ta extended film with a thickness of 6 nm.

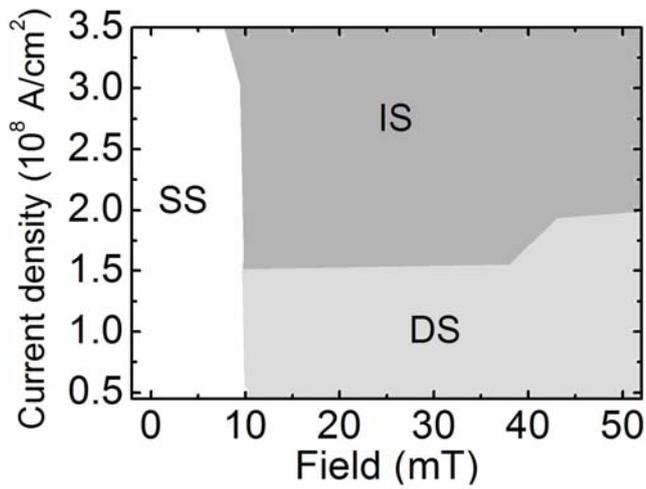

FIG. 2 Phase diagram indicating the switching processes (reversal from positive to negative z-component of magnetization) (direct switching (*DS*), switching via an intermediate state (*IS*), and stochastic switching (*SS*)) and for which range of field and current density they occurs.

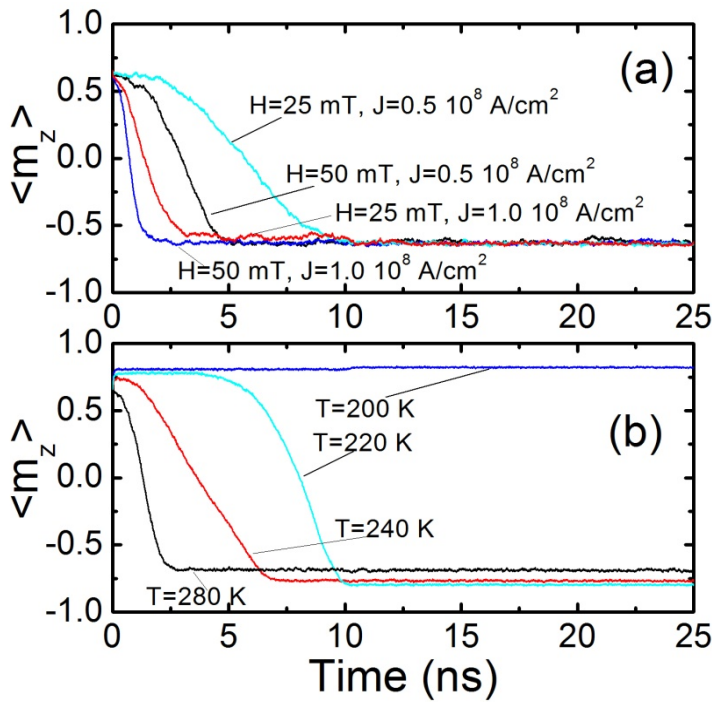

FIG. 3 (a) Time domain traces (normalized average z-component of the magnetization) for the direct switching processes achieved at different field and current density as indicated in the figure and fixed temperature (300 K). (b) Time domain traces (normalized average z-component of the magnetization) ($H$=50 mT $J$=1.0x10$^8$ A/cm$^2$) for different values of temperature as indicated in the figure.

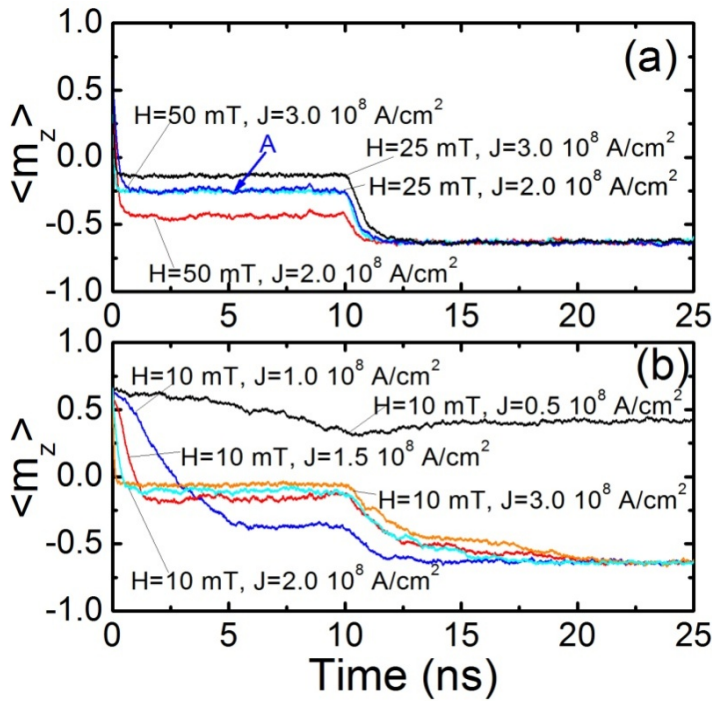

FIG. 4 Time domain traces (normalized average z-component of the magnetization) for the (a) switching processes achieved via an intermediate state and (b) stochastic switching. The different field and current density of the simulations are indicated in the figure, the temperature is fixed (300 K).

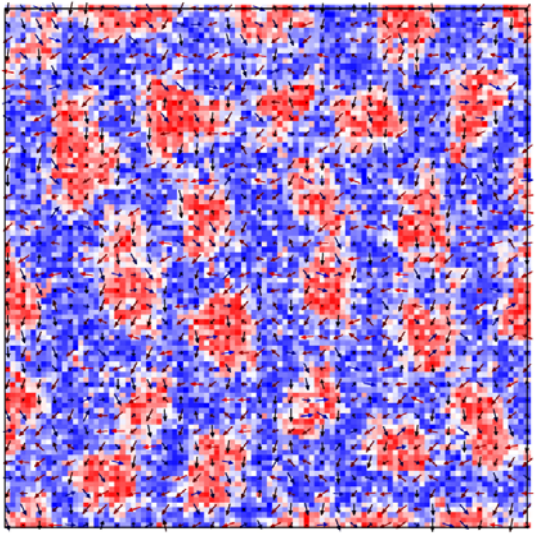

FIG. 5 Example of spatial distribution of the magnetization in the intermediate state ($H=25$mT, $J=2.0 \times 10^8$ A/cm$^2$ point A) achieved before the current pulse is switched off (the color is related to the z-component of the magnetization blue positive red negative, the arrows indicate the in-plane component of the magnetization).

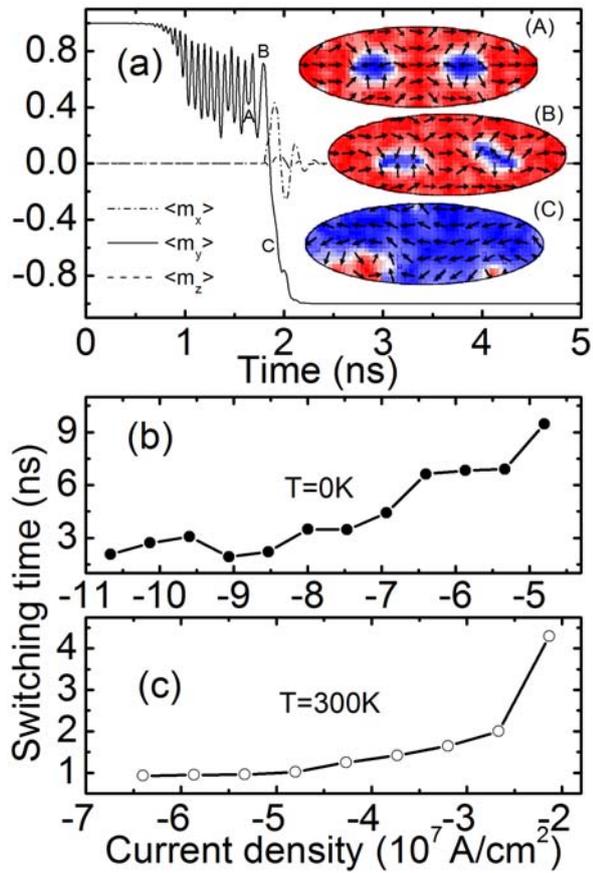

FIG. 6: (a) Example of time evolution of the average normalized magnetization ($x$, $y$ and $z$-component) for the switching process from parallel (P) to antiparallel (AP) achieved for $H=0$ mT and $J=9\times10^7$A/cm$^2$. (b) and (c) switching time without and with thermal fluctuations ($T=300$ K) averaged over 50 iterations.

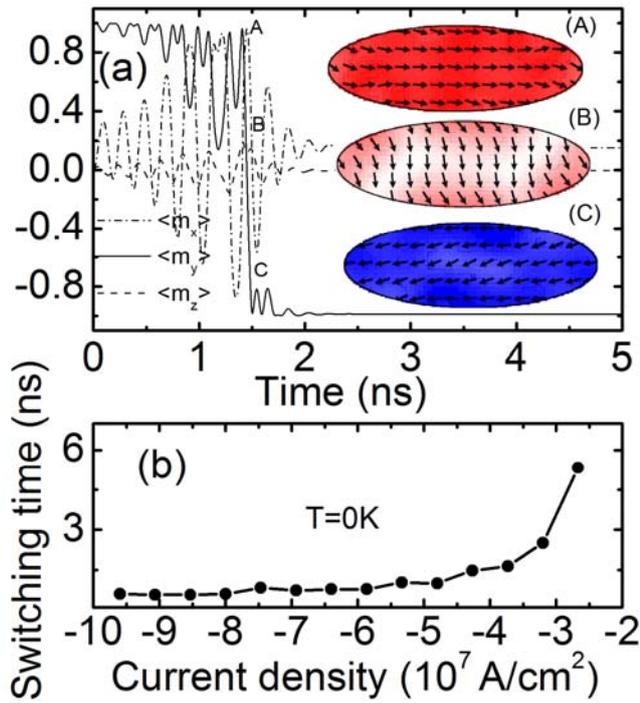

FIG. 7 (a) Example of time evolution of the average normalized magnetization ($x$, $y$ and $z$-component) for the switching process from parallel (P) to antiparallel (AP) achieved for $H=5$ mT and $J=4.25\times10^7$A/cm$^2$. (b) switching time as function of the current density for $H=5$ mT and $T=0$ K.